# Strongly Nonlinear Topological Phases of Cascaded Topoelectrical Circuits

Jijie Tang,[1,*] Fangyuan Ma,[2,*] Feng Li,[2†] Honglian Guo,[1†] Di Zhou[2,†]

[1]College of Science, Minzu University of China, Beijing 100081, China

[2]Centre for Quantum Physics, Key Laboratory of Advanced Optoelectronic Quantum Architecture and Measurement (MOE), School of Physics, Beijing Institute of Technology, Beijing 100081, China

*These authors contributed equally to this work.

†Corresponding author: phlifeng@bit.edu.cn, hlguo@muc.edu.cn, dizhou@bit.edu.cn

**Abstract:**

Circuits provide ideal platforms of topological phases and matter, yet the study of topological circuits in the strongly nonlinear regime, has been lacking. We propose and experimentally demonstrate strongly nonlinear topological phases and transitions in one-dimensional electrical circuits composed of nonlinear capacitors. Nonlinear topological interface modes arise on domain walls of the circuit lattices, whose topological phases are controlled by the amplitudes of nonlinear voltage waves. Experimentally measured topological transition amplitudes are in good agreement with those derived from nonlinear topological band theory. Our prototype paves the way towards flexible metamaterials with amplitude-controlled rich topological phases and is readily extendable to two and three-dimensional systems that allow novel applications.

## Introduction

Topological phases of matter have been widely studied in different areas of physics, such as photonic [1-11], acoustic [12-20], mechanical [21-29], plasmonic [30, 31] and electrical circuit systems [32-45]. Most of the studies of topological systems are limited to the linear regime. Current advances combine topology with weak nonlinearity and give rise to exotic properties, such as topological solitons [46-48], amplitude-controlled topological phase transitions [49-51], non-reciprocal phase transition [52], and frequency conversion of topological modes [53-55], etc., among which the circuit system serves as the ideal platform for exploring properties when topology meets nonlinearity. To date, self-induced topological edge modes and enhanced harmonic generation have been realized in circuits composed of weakly nonlinear elements [50]. However, as perturbative analysis is invalid in the strongly nonlinear regime, the study and application of topological metamaterials with strongly nonlinear interactions, are largely limited. Recently, Berry phase of strongly nonlinear dynamics has been established, which extends the topological bulk-boundary correspondence to the strongly nonlinear regime from the theoretical perspective [56]. However, the experimental observations and characterization of the strongly nonlinear topological physics are still demanding.

In this work, we experimentally demonstrate the strongly nonlinear topological phases via the observation of topological interface excitations in a cascaded circuit. Reflection symmetry quantizes nonlinear Berry phase, whose topologically non-trivial and trivial integer values are controlled by the amplitudes of the voltage fields and correspond to the emergence and absence of nonlinear topological modes on the interface of two semi-ladders. These highly adjustable electrical circuits and flexible phases open the door to smart, tunable and adaptive topological metamaterials.

*The model*---The considered nonlinear topological model is a one-dimensional circuit under periodic boundary condition (PBC). As schematically shown by Fig.1(a), the diatomic unit cell consists of two identical LC resonators, whose inductance and capacitance are $L = 1\mu H$, and $C = 22pF$, respectively. The other ends of the resonators are grounded such that the functionalities are the analog of mechanical oscillators in elastic networks [57-60]. The resonators are connected by nonlinear voltage-dependent capacitors $C_1(V)$ and linear capacitors $C_2 = 25\ pF$ that serve as the inter-cell nonlinear couplings and intra-cell linear couplings, respectively. Here, the nonlinear capacitors $C_1(V)$ are made up of two varactor diodes that yield mirror symmetry [53], and the voltage dependence is obtained by detecting the resonant frequency of an LC resonator that is constructed from an inductor of $1\mu H$ and the nonlinear capacitor $C_1(V)$. By changing the bias voltage, the corresponding resonant frequencies are measured by network analyzer (KEYSIGHT 5061B). Derived from the resonant frequencies, the voltage dependence of the capacitor $C_1(V)$ is plotted by blue dots in Fig. 1(c), with the maximum and minimum values $C_{1\max} = 37.7pF$ and $C_{1\min} = 5pF$, respectively. We further fit these experimentally measured data using Gaussian function, as shown by the red curve in Fig. 1(c), for numerical computations of the nonlinear topological phases and transitions. For a cascaded circuit of $N$ unit cells, the Lagrangian reads

$$\mathbf{L}\left(\left\{V_n^{(1)}, V_n^{(2)}\right\}\right) = \sum_{n=1}^{N}\left[\mathbf{L}_n - U_1\left(V_{n-1}^{(2)} - V_n^{(1)}\right) - U_2\left(V_n^{(1)} - V_n^{(2)}\right)\right], \quad (1)$$

where $\mathbf{L}_n$ is the total Lagrangian of the two LC resonators of the $n$th unit cell, $U_{j=1,2}$ denote the potential energies of the nonlinear and linear capacitors, and $V_n^{(j=1,2)}$ are the two voltage fields of the unit cell, as marked in Fig. 1(a). While the potential energy of linear capacitors is harmonic, the energy of nonlinear capacitors $U_1(V) = \int_0^V (V - x) C_1(x)\,dx$ is strongly anharmonic for large biased voltage $V$, which forbids the availability of linear analysis. Due to the intrinsic structural symmetry of the nonlinear capacitor and the ladder circuit, the Lagrangian stays invariant under reflection transformation

$$\mathbf{L}\left(\left\{V_n^{(1)}, V_n^{(2)}\right\}\right) = \mathbf{L}\left(\left\{V_{-n}^{(2)}, V_{-n}^{(1)}\right\}\right). \quad (2)$$

As we show below, reflection symmetry of the lattice Lagrangian fundamentally quantizes the nonlinear Berry phase, whose non-trivial integer value guarantees the emergence of nonlinear topological interface modes.

The nonlinear dynamics of the considered circuit follow from the Lagrangian equations of motion, which are expressed by four-field generalized nonlinear Schrödinger equations [61]. Spatially repetitive structures enjoy the nonlinear extension of Bloch theorem [56, 61], whose spatial-temporal periodic voltage oscillations take the format of plane-wave nonlinear normal modes [62] $V_k = \left(V_k^{(1)}(\omega t - nk), V_k^{(2)}(\omega t - nk + \phi_k)\right)$. Here, $\omega$ and $k$ are the frequency and wave number respectively, $V_k^{(j=1,2)}(\theta)$ are $2\pi$-periodic wave components, and $\phi_k$ characterizes the relative phase between these two wave components. The frequencies of plane-wave nonlinear normal modes $\omega = \omega(k, A)$ are controlled both by wavenumber $k$ and mode amplitudes $A$, which naturally deviate from their linear counterparts as nonlinearity grows.

Nonlinear normal modes yield reflection symmetry in Eq. (2), from which Berry phase of nonlinear normal modes are guaranteed to pick quantized values (see Ref. [56] and Supplementary Information [61] for details),

$$\gamma(A) = \oint_{\mathrm{BZ}} dk \frac{\sum_l l \left| v_{l,k}^{(2)} \right|^2 \partial_k \phi_k + i \sum_j v_{l,k}^{(j)*} \partial_k v_k^{(j)}}{\sum_{l',j'} l' \left| v_{l',k}^{(j')} \right|^2} = n\pi, \quad n = 0 \text{ or } 1. \tag{3}$$

Here, $A$ stands for the amplitude of the nonlinear voltage modes, and $v_{l,k}^{(j)}(\theta) = (2\pi)^{-1} \int_0^{2\pi} e^{il\theta} V_k^{(j)} d\theta$ is the $l$-th Fourier component of $V_k^{(j)}$. This quantized geometric phase serves as the topological index of the nonlinear circuit dynamics, where $\gamma = \pi$ and $\gamma = 0$ indicate topologically nontrivial and trivial phases, respectively. Upon the increase of mode amplitudes $A$, $\gamma(A)$ cannot change continuously from $\pi$ to 0 due to its topological nature. Nevertheless, it experiences abrupt jumps between distinct integer multiples of $\pi$ as the nonlinear bandgap closes and reopens at the topological transition amplitude $A_c$. This nonlinear topological transition can be intuitively understood by referring to the transitions of linear SSH circuit in Fig. 1(b), whose intercell and intracell couplings are $C_3$ and $C_2$, respectively. When $C_2 < C_3$, the intracell coupling is weaker ($C_2 > C_3$, the intracell coupling is stronger), the linear topological number is in the non-trivial (trivial) phase. Likewise, the nonlinear Berry phase is in the non-trivial (trivial) phase when $C_2 < C_1(V)$ for weaker intracell coupling ($C_2 > C_1(V)$ for stronger intracell coupling), as we discuss below.

As shown in Fig. 1(d), we numerically compute the nonlinear band gap in the circuit system with the unit cells addressed in Fig. 1(a). The nonlinear band gap experiences topological phase transition as voltage amplitudes rise. In the linear regime, the initial bandgap opens, and the topological number $\gamma(A = 0) = \pi$ indicates that the circuit system is in the non-trivial phase. As voltage amplitudes rise, topological invariance states that quantized nonlinear Berry phase should stay unchanged as $\gamma(A < A_c) = \pi$, provided that the nonlinear band gap remains open, where $A_c = 2.97\ V$ is the topological phase transition amplitude. The nonlinear gap closes at this critical amplitude $A_c$, as pictorially depicted by the vanishing gap in Fig. 1(d), where nonlinear Berry phase becomes ill-defined. We define the degree of nonlinearity $\frac{C_1(0) - C_1(A)}{C + C_1(0) + C_2}$ by comparing the nonlinear part of $C_1(A)$ and the linear part of all capacitors $C, C_1(0)$ and $C_2$. At the transition amplitude $A_c$, the degree of nonlinearity reads 0.332, which demonstrates the strongly nonlinear regime of the underlying circuit dynamics [56, 62]. The bandgap reopens above $A_c$ (Fig. 1(d), whose topological index is well-defined again to pick the integer value $\gamma(A > A_c) = 0$ in the trivial phase.

Fig. 2 addresses both the theoretical and experimental transition amplitudes of the topological index in the strongly nonlinear circuit dynamics, as we treat the linear capacitor $C_2$ as the varying parameter in the horizontal axis. As the amplitude of the voltage fields grows, integer-valued topological Berry phase jumps from $\gamma = \pi$ to 0, as indicated by the nonlinear topological phase transition of the unit cell structure in Fig. 1(a). The theoretical scenario of the transition amplitude is based on the matching condition of the frequencies $\omega(k = \pi, \phi_\pi = 0, A_c) = \omega(k = \pi, \phi_\pi = \pi, A_c)$ of the nonlinear normal modes at the time-reversal-invariant-momentum $k = \pi$ with even ($\phi_\pi = 0$) and odd ($\phi_\pi = \pi$) parities. Given the nonlinear capacitor $C_1(V)$ of Fig. 1(a), we theoretically compute a series of topological transition amplitudes $A_c$ by varying the linear capacitor $C_2$, and plot the relationship between transition amplitudes and the linear capacitor using the blue curve in Fig. 2. For example, the transition amplitude for $C_2 = 23pF$ is 3.65 $V$, whose topological transition is captured by the inset of Fig. 2. For $C_2 > C_{1\max} = 37.7\,pF$, the topological phase stays trivial for all voltage amplitudes, and thus the system cannot experience nonlinear topological transition. As $C_2$ drops below $C_{1\max} = 37.7\,pF$, increasing amplitudes are needed to achieve the topological phase transition. Meanwhile, we experimentally probe these transition amplitudes by identifying the emergence of nonlinear topological interface modes. The experimentally measured transition amplitudes for $C_2 = 15pF$, $18pF$, $20pF$, $22pF$, $25pF$, $27pF$, $32pF$ are denoted by square marks in Fig. 2, which qualitatively agree with the aforementioned simulation results. Furthermore, we investigate how the nonlinear topological transition amplitude is affected by fluctuations in the nonlinear capacitors $C_1$ with a range of $\pm 10\%$. In Fig. 2, the lower bound of the blue area indicates that the theoretical curve for the topological transition voltage exhibits better agreement with experimental measurements for a value of $0.9C_1$. Deviations between theory and experiment may also arise from fluctuations in the linear coupling strength of $C_2$, on-site resonators $L$ and $C$, and resistance, which is set to zero in theory but non-zero in experiments.

Based on these numerical demonstrations of topological phases and transitions in the nonlinear circuit model, we experimentally conduct the corresponding nonlinear topological physics in real space. According to the nonlinear extension of bulk-boundary correspondence [63], topological physics can be manifested by the emergence and absence of nonlinear topological modes on the interface of two semi-lattices. To observe the evolution of topological interface modes, we build two prototypes in Fig. 3 and Fig. 4, and experimentally investigate the spatial profile of the impedance along the circuit board in response to external excitation power.

The electrical circuits are built on the Printed Circuit Board, with the tolerance of chip capacitors and chip inductors $\pm 5\%$ and $\pm 10\%$, respectively. We measure the impedance response of circuit system by generating a chirp voltage signal from a function generator (KEYSIGHT 33600A), and subsequently enlarge the signal by a power amplifier (Minicircuit ZHL-6A-S+). The impedance is measured by the frequency response of the voltage and current on the top end of the $LC$ resonators by an oscilloscope (KEYSIGHT DSOX4054A) controlled by a computer. We probe the voltage responses in all unit cells and measure their local impedance by raising the excitation power from $0.038\ V$ to $7.84\ V$, to experimentally measure the responding interface modes.

The first prototype in Fig. 3(a) considers two semi-infinite ladder circuits, whose unit cells are enclosed by the green and red dashed boxes respectively, to construct a mutual interface between them. We encircle the unit cells of the left-sided and right-sided semi-lattices using the green and red dashed boxes, respectively. These gauge choices of the unit cells yield open boundary conditions on both sides of the experimental circuit board in Fig. 3(b). Other unit cell choices of the left semi-lattice may cause problems, because left the open boundary can slice the unit cell at $N = -8$ into half, making it un-defined.

In Fig. 3(b), we experimentally build the circuit board based on the design principle of Fig. 3(a), where both the left and right sides of the interface contain 4 unit cells. On the right side, the unit cells of the semi-lattice are composed of purely linear electrical elements with $C_2 = 25\, pF$ and $C_3 = 37pF$ as the intracell and intercell couplings. The topological number is fixed at $\gamma_{\mathrm{right}} = \pi$, and the linear band gap is marked by the blue dashed box on the right semi-lattice of Figs. 3(c, d, e). On the left side, the intracell and intercell couplings are $C_1(V)$ and $C_2$, respectively. In the weakly nonlinear regime, we approximate $C_1(V = 0.038\ \mathrm{V}) = C_3$, and hence, the topological number of the left semi-lattice is $\gamma_{\mathrm{left}} = 0$ is in line with the linear SSH model. This index is different from the right semi-lattice, because the unit cell choices are different on the two sides of the interface to yield open boundary conditions. As a result, $C_1(V = 0.038\ \mathrm{V}) = C_3, C_2$ and $C_3$, together appear alternatively in real space to constitute a lattice without an interface, and no topological interface modes are expected in the weakly nonlinear regime. As pictorially manifested by the dark band gap that ranges from 15.36MHz to 17.88MHz in Fig. 3(c), topological voltage modes cannot arise on the interface, which is in line with purely linear SSH models. In Fig. 3(d), the interface is on the verge of nonlinear topological phase transition for the external triggering power at $3.09\ V$, which approaches the transition amplitude $A_c = 2.97\ V$, and the left nonlinear band gap closes. As the amplitude further rises to 7.84 V in Fig. 3(e), the intracell coupling of the left semi-lattice, $C_1(V \approx 7.84\ \mathrm{V}) \approx 5\mathrm{pf}$, becomes weaker. The nonlinear band gap reopens above the topological transition amplitude, as depicted by the blue dashed box in the left semi-lattice of Fig. 3(e), leading to the topological numbers $(\gamma_{\mathrm{left}}, \gamma_{\mathrm{right}}) = (\pi, \pi)$ in the large-amplitude regime. Since the band gaps of the left and right semi-lattices mismatch, nonlinear interface modes only arise on the right semi-lattice (left semi-lattice) within the frequency range between 15.36 MHz and 17.88 MHz (between 17.88 MHz and 19.70 MHz) as the same frequency is in the conducting band on the other side of the interface, which enables bulk mode excitations. These experimental results can be verified using LTspice simulations in Figs. 3(f, g, h), where the nonlinear capacitors $C_1(V)$ are numerically replaced by purely linear ones of $37pF$ in Fig. 3(c), $25pF$ in Fig. 3(d), and $15pF$ in Fig. 3(e), respectively.

In the second prototype, namely Fig. 4(a), the interface is composed by two semi-ladder circuits that are mirror-reflection of one another. Following the unit cell convention in Fig. 3, the unit cells in Fig. 4 are encircled by the left and right green dashed boxes. This gauge choice is not only compatible with open boundary conditions, but also yields mirror symmetry regarding the interface. Given that the external power is $0.065\ V$ in Fig. 4(c), stronger capacitors, $C_1(V = 0.065\text{ V}) \approx C_3$, are connected to the interface, whose topological phases of the numbers $(\gamma_{\text{left}}, \gamma_{\text{right}}) = (\pi, \pi)$ are analogous to linear SSH models. We observe the nonlinear topological interface mode, which is also in line with the linear topological interface modes of linear SSH circuits. The topological mode becomes blur in Fig. 4(d) when the external power $1.91\ V$ approaches the critical point $A_c = 2.97\ V$, as indicated by the closure of the nonlinear band gap. As the exciting power further grows to $6.17\ V$ in Fig. 4(e), the interface is in the non-topological regime, which manifests a nonlinear localized mode. This mode is not topological because the frequency can shift into the nonlinear band by tuning the coupling parameters, as indicated by the interface studies of nonlinear topological mechanics [55, 57]. These nonlinear topological physics can be verified by performing numerical simulations in Figs. 4(f, g, h), where the nonlinear capacitors, $C_1(V)$, are now replaced by linear capacitors ($37pF$, $25pF$, and $15pF$) in the calculations of LTspice. It is worth emphasizing that all these nonlinear topological phases, transitions, and interface modes are flexibly controlled by the external input power without entangling/disentangling the hardware of the platform.

In summary, we construct and experimentally demonstrate nonlinear topological modes on the interface of two cascaded semi-ladder electrical circuits. Nonlinear Berry phase is quantized by reflection symmetry of the underlying circuit structure, and guarantees nonlinear topological interface modes in the non-trivial regime. Amplitude-induced topological transitions are naturally manifested from the conversion between topologically non-trivial and trivial interface modes, whose topological transition voltage amplitudes are in good agreement between experimental measurements and simulations from nonlinear topological band theory. Our prototype establishes flexible metamaterials with amplitude-controlled rich topological phases and transitions and are readily extendable to higher dimensional platforms.

**Acknowledgments**

This work is supported by the National Natural Science Foundation of China (Grants No. 12102039, No. 12272040, and No. 12074446).

**Author contributions**

All authors contributed extensively to the work presented in this paper. Jijie Tang, and Feng Li carried out the experiments. Fangyuan Ma, and Di Zhou provided the theory and calculations. Di Zhou, and Feng Li wrote the paper and Supplemental Material.

**Conflict of interest**

The authors declare that they have no conflict of interest.

**References**

[1] M. Hafezi, S. Mittal, J. Fan, A. Migdall, and J. M. Taylor, Imaging topological edge states in silicon photonics. Nat. Photon. **7**, 1001–1005 (2013).

[2] M. S. Kirsch, Y. Zhang, M. Kremer, L. J. Maczewsky, S. K. Ivanov, Y. V. Kartashov, L. Torner, D. Bauer, A. Szameit, and H. Matthias, Nature Physics. **17** 995 (2021).

[3] M. C. Rechtsman, J. M. Zeuner, Y. Plotnik, Y. Lumer, D. Podolsky, F. Dreisow, S. Nolte, M. Segev and A. Szameit, Photonic Floquet topological insulators. Nature. **496**, 196–200 (2013).

[4] H. Schomerus, Topologically protected midgap states in complex photonic lattices. Opt. Lett. **38**, 1912–1914 (2013).

[5] A. B. Khanikaev, S. H. Mousavi, W. K. Tse, M. Kargarian, A. H. MacDonald and G. Shvets, Photonic topological insulators. Nat. Mater. **12**, 233–239 (2013).

[6]. L. Lu, J. D. Joannopoulos, and M. Soljačić, Topological photonics. Nat. Photonics. **8**, 821 (2014).

[7] T. Tuloup, R. W. Bomantara, C. H. Lee, and J. Gong, Nonlinearity induced topological physics in momentum space and real space, Physical Review B. **102** 115411 (2020).

[8] R. W. Bomantara, and W. Zhao, L. Zhou, and J. Gong, Nonlinear Dirac cones, Physical Review B. **96** 121406 (2017).

[9] J. M. Zeuner, M. C. Rechtsman, Y. Plotnik, Y. Lumer, S. Nolte, M. S. Rudner, M. Segev and A. Szameit, Observation of topological transition in the bulk of a non-Hermitian system. Phys. Rev. Lett. **115**, 040402 (2015).

[10] J. Jiang, J. Ren, Z. Guo, W. Zhu, Y. Long, H. Jiang, H. Chen. Seeing topological winding number and band inversion in photonic dimer chain of split-ring resonators. Phys Rev B. **101**, 165427 (2020).

[11] Z. Guo, J. Jiang, H. Jiang, J. Ren, H. Chen. Observation of topological bound states in a double Su-Schrieffer-Heeger chain composed of split ring resonators. Phys. Rev. Research. **3**, 013122 (2021).

[12] Z. Yang, F. Gao, X. Shi, X. Lin, Z. Gao, Y. Chong, and B. Zhang, Topological acoustics. Phys. Rev. Lett. **114**, 114301 (2015).

[13] A. Souslov, Z. Van, C. Benjamin, D. Bartolo, and V. Vitelli, Topological sound in active-liquid metamaterials, Nature Physics **13** 1091 (2017).

[14] G. Lee and J. Rho, Piezoelectric energy harvesting using mechanical metamaterials and phononic crystals, Communications Physics **5,** 94 (2022).

[15] R. Susstrunk and S. D. Huber, Observation of phononic helical edge states in a mechanical topological insulator, Science, **349** 47 (2015).

[16] C. He, X. Ni, H. Ge, X. Sun, Y. Chen, M. Lu, X. Liu and Y. Chen, Acoustic topological insulator and robust one-way sound transport. Nat. Phys. **12**, 1124 (2016).

[17] V. Peano, C. Brendel, M. Schmidt, and F. Marquardt, Topological phases of sound and light. Phys. Rev. X **5**, 031011 (2015).

[18] M. Xiao, G. Ma, Z. Yang, P. Sheng, Z. Q. Zhang and C. T. Chan, Geometric phase and band inversion in periodic acoustic system. Nat. Phys. **11**, 240–244 (2015).

[19] H. He, C. Qiu, L. Ye, X. Cai, X. Fan, M. Ke, F. Zhang and Z. Liu. Topological negative refraction of surface acoustic waves in a Weyl phononic crystal. Nature. **560,**

7716 (2018).

[20] J. Lu, C. Qiu, L. Ye, X. Fan, M. Ke, F. Zhang and Z. Liu. Observation of topological valley transport of sound in sonic crystals. Nature Phys **13**, 369–374 (2017).

[21] C. L. Kane, and T. C. Lubensky, Topological boundary modes in isostatic lattices, Nature Physics, **10**, 39, (2014).

[22] J. Paulose, B. G. Chen, and V. Vitelli, Topological modes bound to dislocations in mechanical metamaterials, Nature Physics **11** 153 (2015).

[23] H. Xiu, H. Liu, A. Poli, G. Wan, K. Sun, E. M. Arruda, and X. Mao, Topological transformability and reprogrammability of multistable mechanical metamaterials, Proceedings of the National Academy of Science (2022).

[24] D. Zhou, L. Zhang and X. Mao, Topological edge floppy modes in disordered fiber networks, Phys. Rev. Lett. **120**, 068003 (2018).

[25] M. Fruchart and V. Vitelli, Symmetries and dualities in the theory of elasticity, Phys. Rev. Lett. **124** 248001 (2020).

[26] J. Ma, D. Zhou, K. Sun, X. Mao, and S. Gonella, Edge Modes and Asymmetric Wave Transport in Topological Lattices: Experimental Characterization at Finite Frequencies, Phys. Rev. Lett. **121**, 094301 (2018).

[27] H. Liu, D. Zhou, L. Zhang, D. K. Lubensky, and X. Mao, Topological floppy modes in models of epithelial tissues, Soft Matter **17** 8624 (2021).

[28] M. Rosa, and M. Ruzzene, and E. Prodan, Topological gaps by twisting, Communications Physics, **4** 1 (2021).

[29] D. Zhou, L. Zhang and X. Mao, Topological boundary floppy modes in quasicrystals, Phys. Rev. X, **9**, 021054 (2019).

[30] Y. Fu, and H. Qin, Topological phases and bulk-edge correspondence of magnetized cold plasmas, Nature Communications, **12**, 1, (2021).

[31] Y. Fu, and H. Qin, The dispersion and propagation of topological Langmuir-cyclotron waves in cold magnetized plasmas, Journal of Plasma Physics, **88**, 4, (2022).

[32]. V. V. Albert, L. I. Glazman, and L. Jiang, Topological properties of linear circuit lattices. Phys. Rev. Lett. **114**, 173902 (2015).

[33] S. Imhof, C. Berger, F. Bayer, J. Brehm, L. W. Molenkamp, T. Kiessling, F.

Schindler, C. H. Lee, M. Greiter, T. Neupert and R. Thomale, Topolectrical circuit realization of topological corner modes. Nat. Phys. **14**, 925–929 (2018).

[34]. N. Jia, C. Owens, A. Sommer, D. Schuster, and J. Simon, Time- and site-resolved dynamics in a topological circuit. Phys. Rev. X **5**, 021031 (2015).

[35] T. Goren, K. Plekhanov, F. Appas, and K. L. Hur, Topological Zak phase in strongly coupled LC circuits. Phys. Rev. B **97**, 041106(R) (2018).

[36] W. Zhu, S. Hou, Y. Long, H. Chen, and J. Ren, Simulating quantum spin Hall effect in the topological Lieb lattice of a linear circuit network. Phys. Rev. B **97**,075310 (2018).

[37]. M. Serra-Garcia, R. Süsstrunk. And S. D. Huber, Observation of quadrupole transitions and edge mode topology in an LC network. Phys. Rev. B **99**, 020304(R) (2019).

[38] T. Helbig, T. Hofmann, et al. Generalized bulk–boundary correspondence in non-Hermitian topolectrical circuits. Nat. Phys. **16** 747–750 (2020).

[39] C. H. Lee, S. Imhof, C. Berger, F. Bayer, J. Brehm, L. W. Molenkamp, T. Kiessling and R. Thomale, Topolectrical circuits. Communications Physics. 1(1): 39 (2018).

[40] D. Zou, T. Chen, W. He, J. Bao, C. H. Lee, H. Sun, Observation of hybrid higher-order skin-topological effect in non-Hermitian topolectrical circuits. Nature Communications. **12**(1): 7201 (2021).

[41] T. Hofmann, T. Helbig, F. Schindler, N. Salgo, M. Brzezińska, M. Greiter, T. Kiessling, D. Wolf, A. Vollhardt, A. Kabaši, C. H. Lee, A. Bilušić, R. Thomale, and T. Neupert, Reciprocal skin effect and its realization in a topolectrical circuit. Phys. Rev. Research. **2**(2): 023265 (2020).

[42] H. Yang, Z.-X. Li, Y. Liu, Y. Cao, and P. Yan, Observation of symmetry-protected zero modes in topolectrical circuits[J]. Phys. Rev. Research. **2**(2): 022028 (2020).

[43] T. Hofmann, T. Helbig, C. H. Lee, M. Greiter, and R. Thomale, Chiral voltage propagation and calibration in a topolectrical Chern circuit. Phys. Rev. Lett. **122**(24): 247702 (2019).

[44] L. Xiao, T. Deng, K. Wang, G. Zhu, Z. Wang, W. Yi and Peng Xue, Non-Hermitian bulk–boundary correspondence in quantum dynamics. Nat. Phys. **16**(7): 761-766 (2020).

[45] W. Zhu, Y. Long, H. Chen, J. Ren. Quantum valley Hall effects and spin-valley locking in topological Kane-Mele circuit networks. Physical Review B. **99**, 115410 (2019).

[46] Y. Lumer, Y. Plotnik, M. C. Rechtsman and M. Segev, Self-Localized States in Photonic Topological Insulators. Phys. Rev. Lett. **111**, 243905 (2013).

[47] D. Leykam and Y. D. Chong, Edge solitons in nonlinear-photonic topological insulators. Phys. Rev. Lett. **117**, 143901 (2016).

[48] Y. Lumer, M. C. Rechtsman, Y. Plotnik and M. Segev, Instability of bosonic topological edge states in the presence of interactions. Phys. Rev. A. **94**, 021801 (R) (2016).

[49] Y. Hadad, A. B. Khanikaev and A. Alu. Self-induced topological transitions and edge states supported by nonlinear staggered potentials. Phys. Rev. B. **93**, 155112 (2016).

[50]. Y. Hadad, J. C. Soric, A. B. Khanikaev, and A. Alù, Self-induced topological protection in nonlinear circuit arrays. Nat. Electron. **1**, 178 (2018).

[51] H. Xiu, I. Frankel, H. Liu, K. Qian, S. Sarkar, B. C. Macnider, Z. Chen, N. Boechler, and X. Mao, Synthetically Non-Hermitian Nonlinear Wave-like Behavior in a Topological Mechanical Metamaterial, arXiv:2207.09273.

[52] M. Fruchart, R. Hanai, P. B. Littlewood, and V. Vitelli, Non-reciprocal phase transitions, Nature **592** 363 (2021).

[53] Y. Wang, L-J. Lang, C. H. Lee, B. Zhang and Y. D. Chong. Topologically enhanced harmonic generation in a nonlinear transmission line metamaterial. Nat. Commun. **10**, 1102 (2019).

[54] D. Zhou, J. Ma, K. Sun, S. Gonella, and X. Mao, Switchable phonon diodes using nonlinear topological Maxwell lattices, Phys. Rev. B, **101**, 104106 (2020).

[55] R. K. Pal, J. Vila, M. Leamy, and M. Ruzzene, Amplitude-dependent topological edge states in nonlinear phononic lattices. Phys. Rev. E. **97**, 032209 (2018).

[56] D. Zhou, D. Z. Rocklin, M. J. Leamy, and Y. Yao, Topological invariant and anomalous edge modes of strongly nonlinear systems. Nat. Commun. **13**, 3379 (2022).

[57] J. R. Tempelman, K. H. Matlack, and A. F. Vakakis, Topological protection in a

strongly nonlinear interface lattice. Phys. Rev. B. **104**, 174306 (2021).

[58] J. Vila, G. Paulino, and M. Ruzzene, Role of nonlinearities in topological protection: Testing magnetically coupled fidget spinners, Phys. Rev. B **99** 125116 (2019).

[59] D. Zhou and J. Zhang, Non-Hermitian topological metamaterials with odd elasticity, Phys. Rev. Research, **2**, 023173 (2020).

[60] W. Cheng and G. Hu, Acoustic skin effect with non-reciprocal Willis materials, Appl. Phys. Lett. **121**, 041701 (2022).

[61] See Supplementary Information for experimental setup and measurement, the nonlinear topological band theory, nonlinear Berry phase, and topological phase transitions.

[62] A. F. Vakakis, L. I. Manevitch, Y. V. Mikhlin, V. N. Pilipchuk, and A. A. Zevin, Normal modes and localization in nonlinear systems. (Springer, 2001).

[63] C. Shang, Y. Zheng, and B. A. Malomed, Weyl solitons in three-dimensional optical lattices, Phys. Rev. A **97** 043602 (2018).

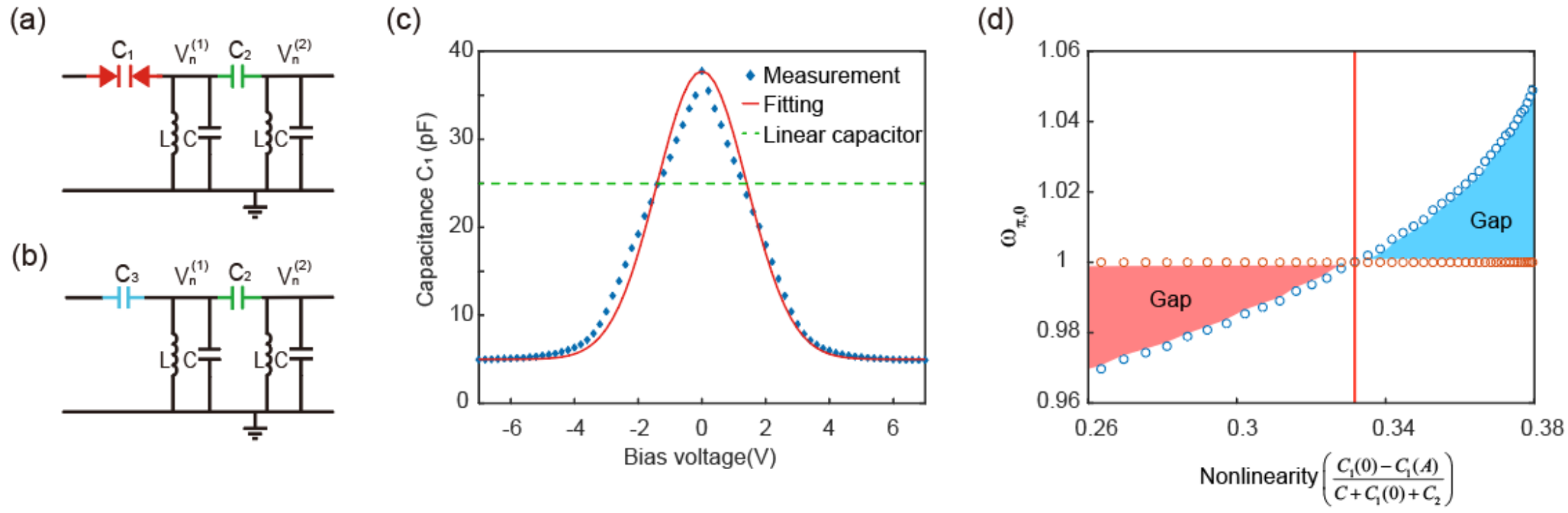

**Fig. 1.** (a) The unit cell of the underlying nonlinear topoelectrical circuit. (b) The unit cell of the linear circuit that helps to construct an interface between two semi-lattices in Fig. 3. (c) The voltage-dependence of the nonlinear capacitors, where the dots represent the data measured by the network analyzer, and the fitting Gaussian curve is adopted for numerical computations. (d) The numerical calculation of the nonlinear band gap and topological phase transitions of the circuit unit cell in (a) for growing nonlinearity as voltage amplitude increases. At the transition point marked by the vertical red line, the degree of nonlinearity (horizontal axis) reads 0.332, which grants the considered topoelectrical circuit the strongly nonlinear regime.

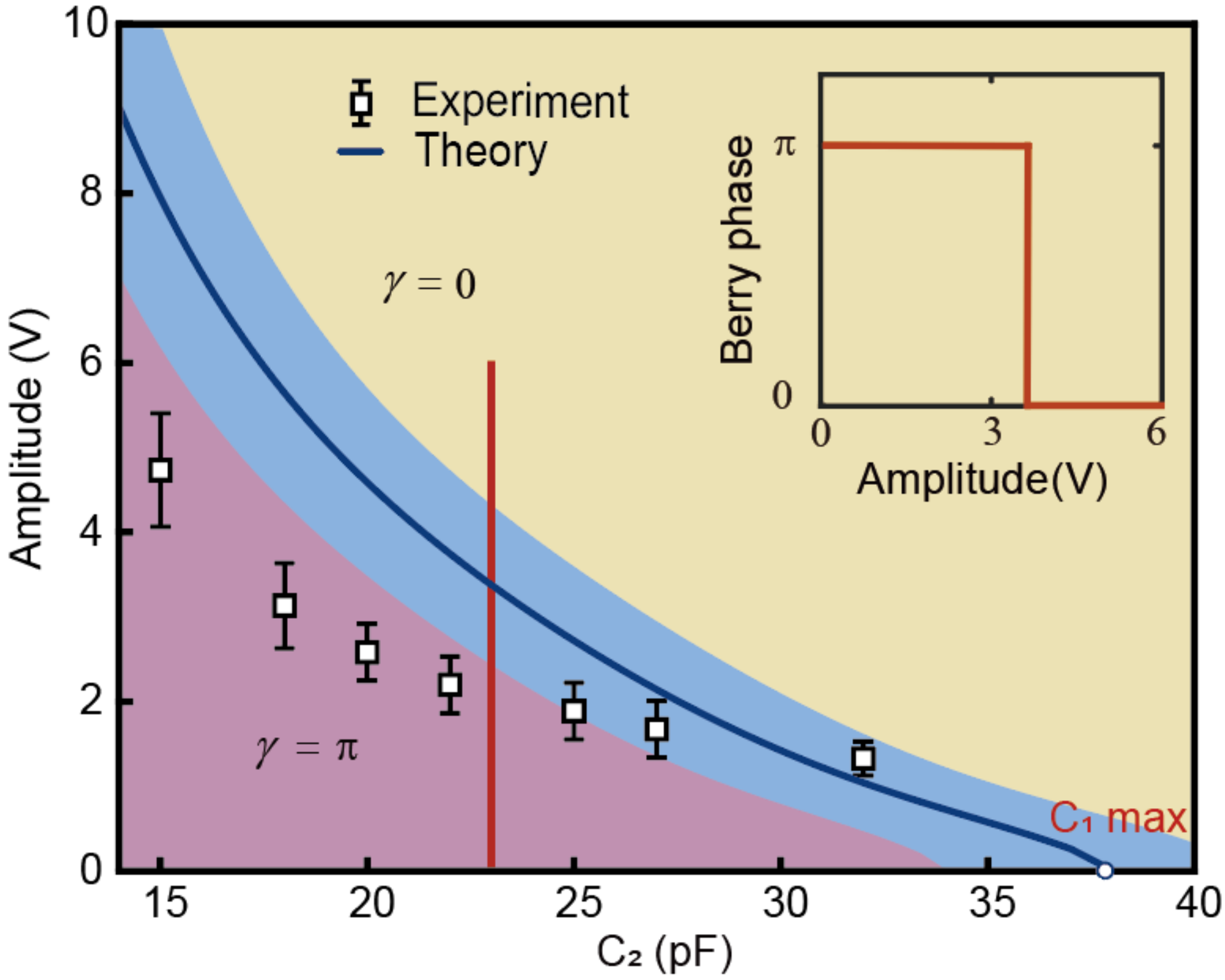

**Fig. 2.** Topological phase diagram of the nonlinear electrical circuit in Fig. 1(a). In the horizontal axis, we vary the parameter of the linear capacitor $C_2$ from 14.0 $pF$ to 37.0 $pF$. The vertical axis represents the amplitude of responding voltage fields. The blue curve denotes the numerical result of the topological transition voltage amplitudes for varying linear capacitor $C_2$ . The blue area depicts the numerical result of the topological transition voltage under the influence of fluctuations in the nonlinear capacitors $C_1$ with a range of $\pm 10\%$. The experimentally measured transition voltages are depicted by the square marks with error bars. The inset illustrates the transition of the topological index when $C_2 = 23pF$.

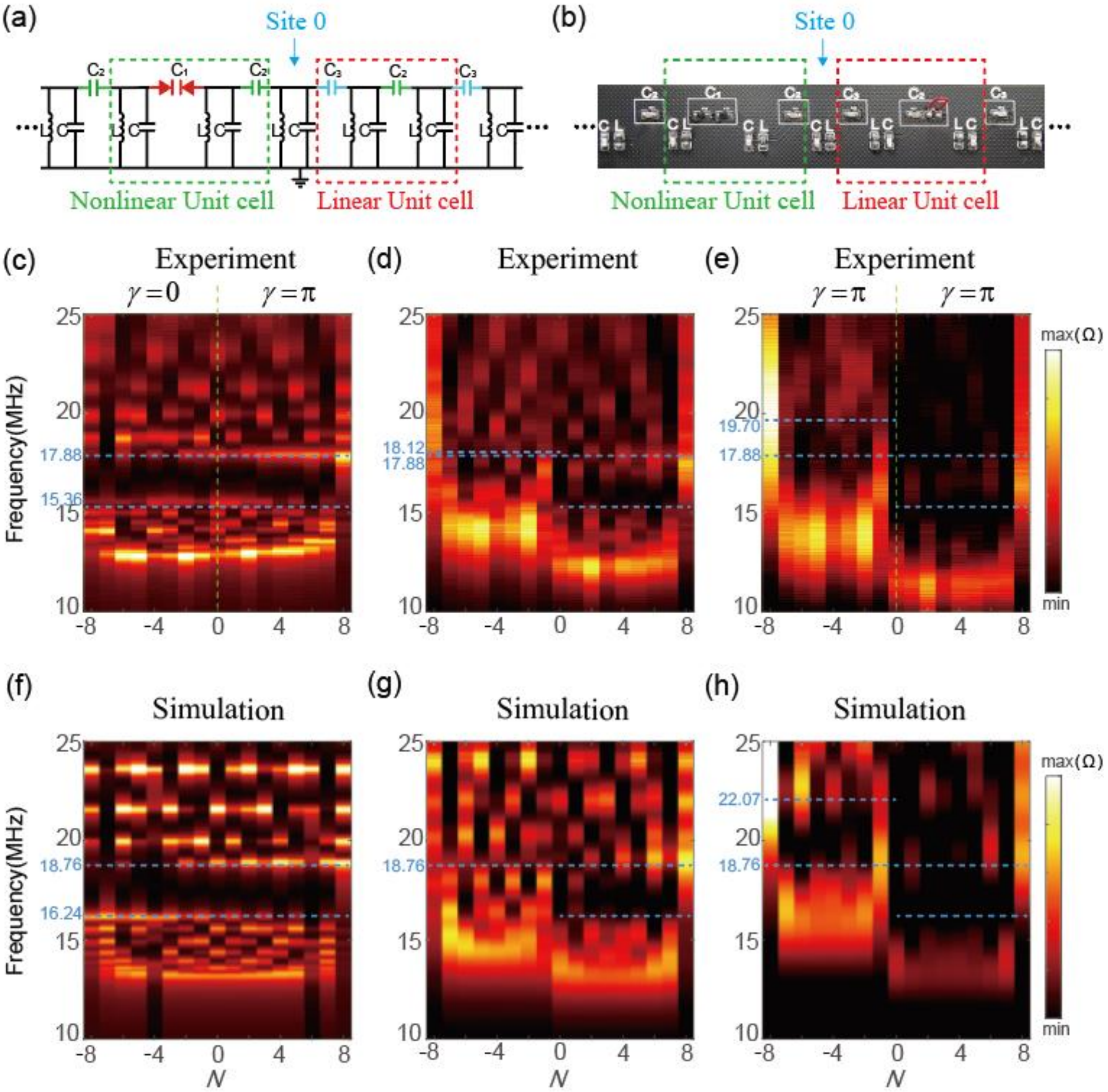

**Fig. 3.** (a) Schematic illustration of the first prototype of the nonlinear topoelectrical circuit, whose interface connects the left and right half-lattices. Encircled by the red dashed box, the right-sided unit cells are composed of linear elements, as shown in Fig. 1(b). The left-sided semi-lattice, as enclosed by the green dashed box, is composed of unit cells whose intra-cell and inter-cell couplings are switched comparing with Fig. 1(a). (b) Photograph of the experimentally constructed circuit board from the design in (a), whose right and left sides contain 4 unit cells each. (c) The absence of topological interface mode in the small amplitude regime. (d) Topological interface mode on the verge of nonlinear topological phase transition. the mode amplitude reads $3.09\,V$, which approaches the transition point $A_c = 2.97\,V$. (e) Nonlinear topological interface mode becomes clearer for amplitude at $7.84\,V$. (f) (g) (h) Impedance diagram of variable capacitor replaced by linear capacitor $37pF$ in (f), $25pF$ (g), and $15pF$ in (h).

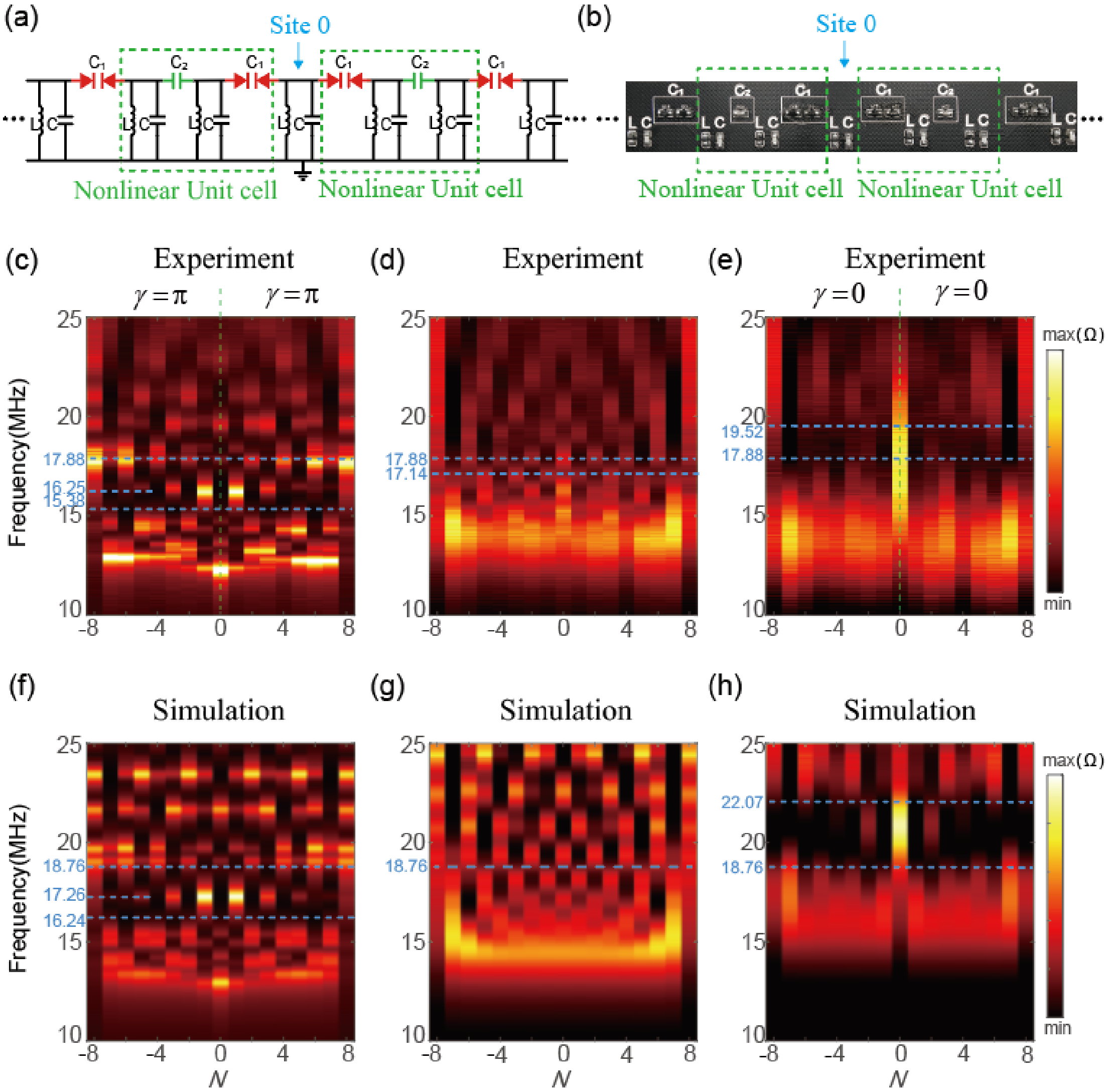

**Fig. 4.** (a) Schematic illustration of the second prototype of the nonlinear topoelectrical circuits. As encircled by the green dashed boxes, the unit cells of the two semi-lattices are designed from Fig. 1(a). (b) Experimental setup of (a). (c) The interface hosts a topological mode with the excitation power of $0.065\ V$. (d) Nonlinear topological interface mode on the verge of disappearance, as the external power $1.91\ V$ approaches the topological transition amplitude $A_c = 2.97\ V$. (e) For the power that further rises to $6.17\ V$, trivial localized modes take place on the interface. (f) (g) (h) Impedance diagram of variable capacitor replaced by linear capacitor $37pF$ in (f), $25pF$ (g), and $15pF$ in (h).

# Strongly Nonlinear Topological Phases of Cascaded Topoelectrical Circuits: Supplementary Information

## I. METHOD

### A. Circuit design

Electrical circuits are built on the printed circuit board (PCB) (thickness, 1.6 mm), with each nonlinear capacitor consisting of a pair of back-to-back varactors (Skyworks Solutions, SMV1255 -011LF). The capacitance ranges from 37.7 pF to 5 pF as the DC voltage changes in the range between 0 to 7V. We use network analyzer and LC series circuit to measure the capacitance curve of the nonlinear capacitor, with the fixed chip capacitors (tolerance, $\pm 5\%$), the fixed chip inductors (tolerance, $\pm 10\%$), and the direct insertion resistors (tolerance, $\pm 1\%$).

### B. Measurement set-up

The experimental observation of the topological interface mode is performed by measuring the impedance of each point marked as $V_n$ in the circuit system. The method of measuring the impedance is the resistance voltage divider circuit method, as shown in Fig.S1 below. The function signal generator produces the chirp signal and adjusts the amplitude. The frequency of the chirp signal ranges from 0Hz to 33.3MHz.

The amplifier (Minicircuit ZHL-6A-S+) amplifies the chirp signal by the factor of 24.2 dB. We use the oscilloscope (KEYSIGHT DSOX4054A) to collect the data of $V_0$ and $V_n$, as shown by the nodes on the two sides of $R_0$ in Fig. S1. Fourier transformation is performed to obtain the frequency spectrum of the response. The impedance $Z_n$ of each node in the circuit is computed via $Z_n = R_0 \cdot V_n/(V_0 - V_n)$.

## II. EQUATIONS OF MOTION OF THE NONLINEAR TOPOELECTRICAL CIRCUIT

In this section, we establish the equations of motion of the cascaded circuit, which is subjected to periodic boundary condition (PBC), and the unit cell is depicted by Fig. S2.

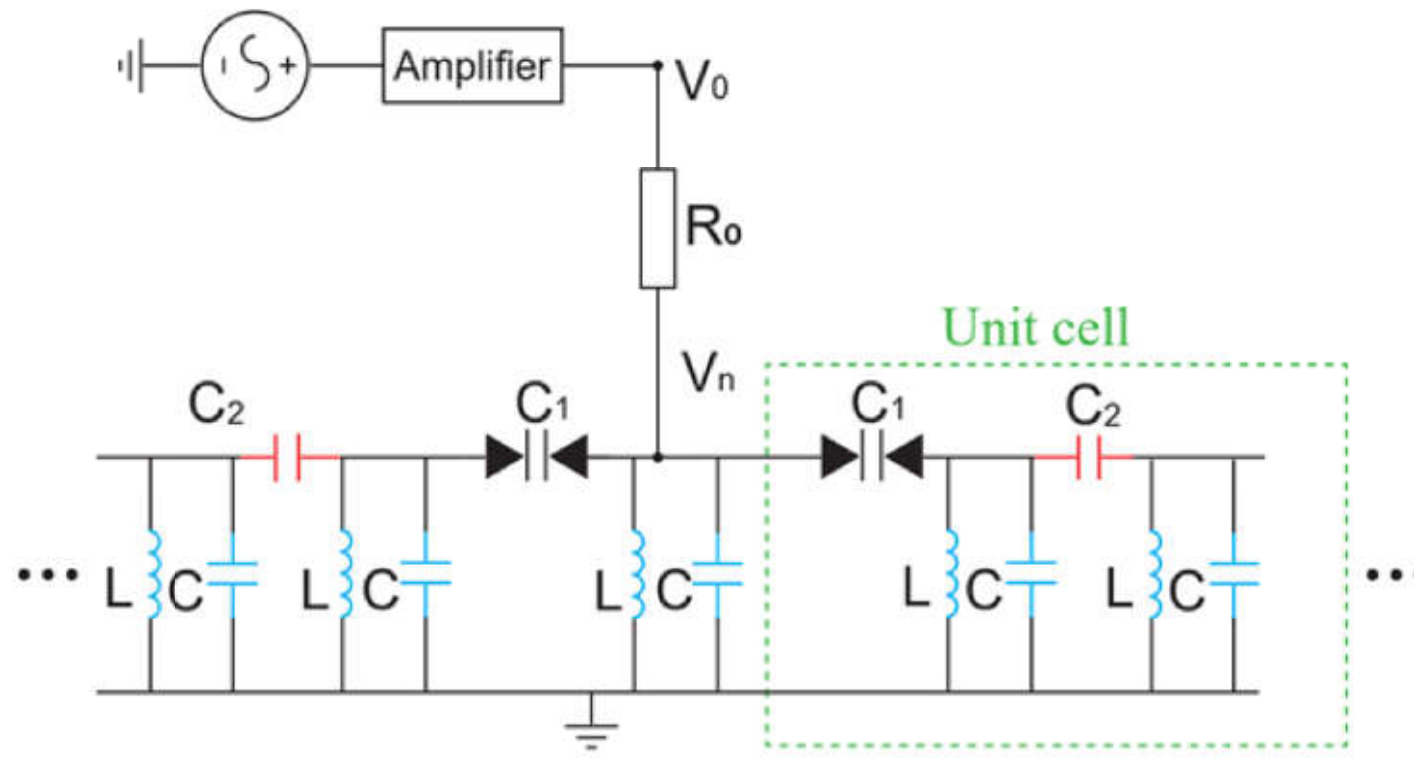

FIG. S1. The method of impedance measurement.

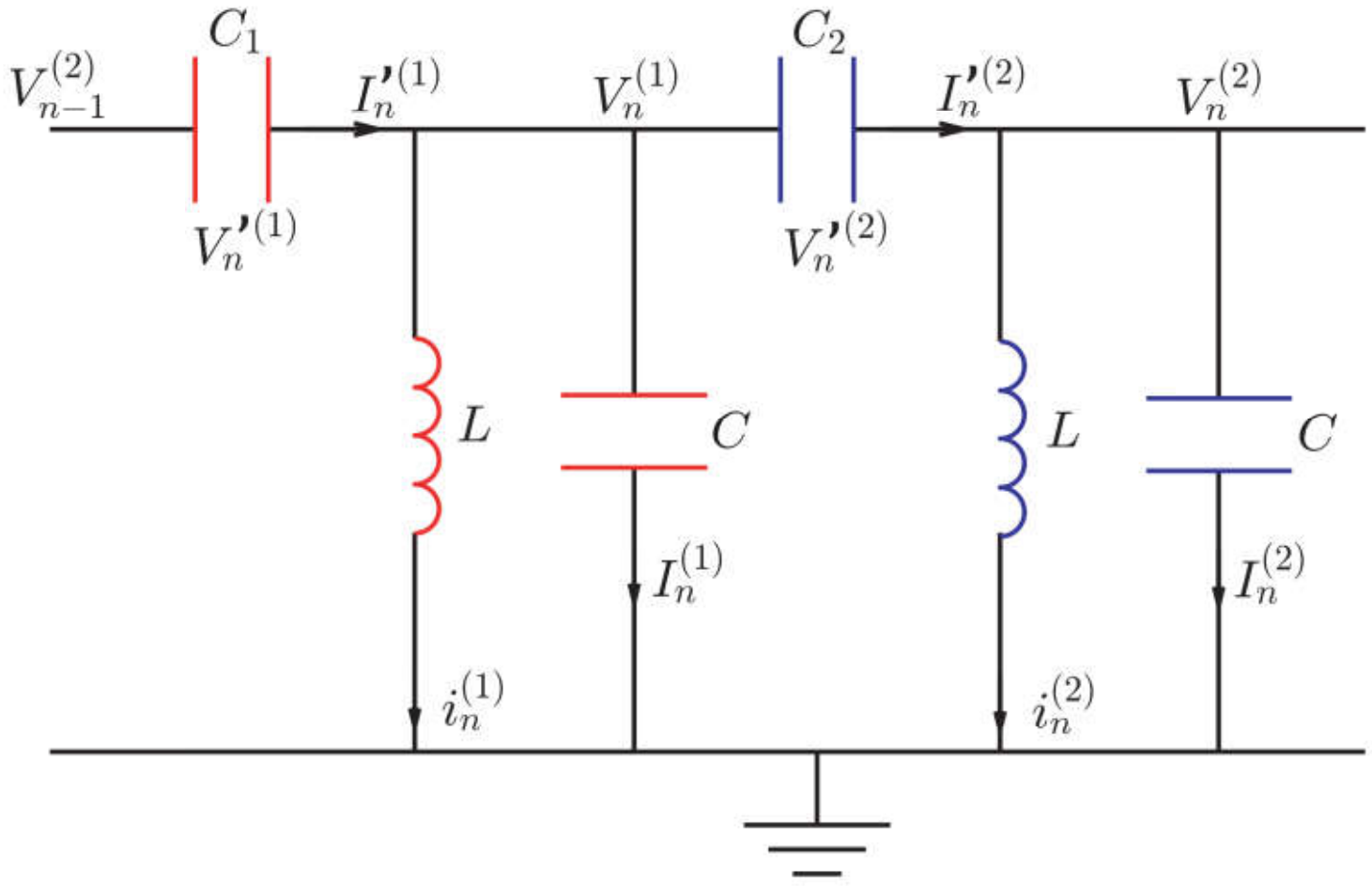

FIG. S2. The unit cell of the nonlinear topoelectrical circuit, where $L$, $C$ are purely linear inductor and capacitor, respectively, and $C_1(V)$, $C_2(V)$ are nonlinear capacitors that vary upon the change of the biased voltages.

We conduct the system Lagrangian, from which the nonlinear motion equations are derived. The circuit Lagrangian reads

$$\mathrm{L}(\{V_n^{(1)}, V_n^{(2)}\}) = \sum_n \left[ \mathrm{L}_n - U_1(V_{n-1}^{(2)} - V_n^{(1)}) - U_2(V_n^{(1)} - V_n^{(2)}) \right], \tag{S1}$$

where $V_n^{(1)}$ and $V_n^{(2)}$ denote the voltage fields in the unit cell, $i_n^{(1)}$ and $i_n^{(2)}$ are the currents that flow in the inductances, $\mathrm{L}_n = \frac{1}{2}L(i_n^{(1)2} + i_n^{(2)2}) - \frac{1}{2}C(V_n^{(1)2} + V_n^{(2)2})$ is the Lagrangian of the two LC resonators in the unit cell, and

$$U_{j=1,2}(V) = \int_0^V (V - x) C_{j=1,2}(x) \tag{S2}$$

are the strongly an-harmonic potential energy of the nonlinear capacitors $C_{j=1,2}(V)$.

It is worth emphasizing that the nonlinear capacitors yield even function $C_{j=1,2}(V) = C_{j=1,2}(-V)$ of the biased voltage, because the nonlinear capacitor is composed of back-to-back varactor diodes that intrinsically yields the structural reflection symmetry [1]. Thus, it is straightforward to arrive at the result that the nonlinear potential energy of the capacitor $C_{j=1,2}$ is an even function of the biased voltage, $U_{j=1,2}(V) = U_{j=1,2}(-V)$. In the next section, we will show that together with the diatomic structure of the unit cells, the nonlinear dynamics of the circuit system yields reflection symmetry, which subsequently quantizes the nonlinear Berry phase.

According to Kirchhoff law, we write the currents of the inductances in terms of the currents that flow in the capacitors. Due to the conservation law of the current at every node, we have the relationships

$$\begin{aligned} i_n^{(1)} &= I_n^{'(1)} - I_n^{(1)} - I_n^{'(2)}, \\ i_n^{(2)} &= I_n^{'(2)} - I_n^{(2)} - I_{n+1}^{'(1)}. \end{aligned} \tag{S3}$$

Then, the currents in the linear capacitors of the LC resonators are given by

$$I_n^{(j=1,2)} = C \frac{dV_n^{(j=1,2)}}{dt}, \tag{S4}$$

whereas the currents that flow in the nonlinear capacitors read

$$\begin{aligned} I_n^{'(1)} &= C_1(V_{n-1}^{(2)} - V_n^{(1)}) \left( \frac{dV_{n-1}^{(2)}}{dt} - \frac{dV_n^{(1)}}{dt} \right), \\ I_n^{'(2)} &= C_2(V_n^{(1)} - V_n^{(2)}) \left( \frac{dV_n^{(1)}}{dt} - \frac{dV_n^{(2)}}{dt} \right). \end{aligned} \tag{S5}$$

Summarizing all these Eqs.(S3, S4, S5), now the system Lagrangian can be expressed only in terms of the voltage fields $V_n^{(j=1,2)}$ and their time-derivatives $\dot{V}_n^{(j=1,2)}$.

Having established the system Lagrangian in terms of the voltage fields, we can now express the nonlinear dynamics of the circuit system using Lagrangian equations motion

$$\frac{d}{dt} \frac{\partial \mathrm{L}}{\partial \dot{V}_n^{(j)}} - \frac{\partial \mathrm{L}}{\partial V_n^{(j)}} = 0 \qquad \text{for} \qquad j = 1, 2. \tag{S6}$$

Despite that the nonlinear dynamics of the voltage fields $V_n^{(j=1,2)}$ appear second-order derivative in time, one can convert them as the first-order time derivatives by defining

$$\Psi_n = (\Psi_n^{(1)}, \Psi_n^{(2)}, \Psi_n^{(3)}, \Psi_n^{(4)}) = (V_n^{(1)}, -\mathrm{i}\dot{V}_n^{(1)}, V_n^{(2)}, -\mathrm{i}\dot{V}_n^{(2)}) \tag{S7}$$

as the classical field variable of the $n$th unit cell. By doing so, one converts the circuit dynamics as the four-field generalized nonlinear Schrödinger equations.

## III. NONLINEAR EXTENSION OF BLOCH THEOREM

Spatially repetitive structures enjoy the nice property of discrete translational symmetry, meaning that the system Lagrangian as well as the nonlinear equations of motion are invariant for translational operations. As a result, spatial-temporal periodic modes in the system take the format of plane-wave nonlinear normal modes [2–4]. For example, plane-wave voltage fields of the circuit system yields the format of $V_k(t) = (V_k^{(1)}(\omega t - kn), V_k^{(2)}(\omega t - kn + \phi_k))$, where $k$ and $\omega$ are the wavenumber and frequency, respectively, and $\phi_k$ characterizes the phase difference between these two wave components. Likewise, the plane-wave format of the four-component classical field reads

$$\Psi_k(t) = \\ (\Psi_k^{(1)}(\omega t - kn + \phi_k^{(1)}), \Psi_k^{(2)}(\omega t - kn + \phi_k^{(2)}), \Psi_k^{(3)}(\omega t - kn + \phi_k^{(3)}), \Psi_k^{(4)}(\omega t - kn + \phi_k^{(4)})), \quad \text{(S8)}$$

where $\Psi_k^{j=1,2,3,4}$ are $2\pi$-periodic wave components, and the phase conditions are chosen by asking that $\mathrm{Re}\,\Psi_k^{(j)}(\theta = 0) = \max(\mathrm{Re}\,\Psi_k^{(j)})$ reaches the maximum of the wave component at $\theta = 0$. Thus, $\phi_k^{(j=1,2,3,4)}$ characterize the relative phase of each wave component.

To prove the nonlinear extension of Bloch theorem, we consider a temporal-periodic solution of the nonlinear circuit, denoted as $\Psi_n(\omega t)$, where $\omega$ is the mode frequency. The mode follows the equations of motion in Eq.(S6), and the corresponding system Lagrangian is $\mathrm{L}(\Psi_n(\omega t))$. Due to the discrete translational symmetry of the underlying circuit lattice, it is natural to find a partner nonlinear solution $\tilde{\Psi}_n(\omega t)$, which yields $\tilde{\Psi}_{n+1}(\omega t) = \Psi_n(\omega t)$. $\tilde{\Psi}_n(\omega t)$ also satisfies the nonlinear equations of motion as the system Lagrangian stays invariant via $\mathrm{L}(\tilde{\Psi}_n(\omega t)) = \mathrm{L}(\Psi_n(\omega t))$. As there is no degeneracy of nonlinear normal modes, the nonlinear solutions $\tilde{\Psi}_n(\omega t)$ and $\Psi_n(\omega t)$ have to be the same solution. Therefore, they can only differ by a phase factor $\Delta\theta$, with $\tilde{\Psi}_n(\omega t + \Delta\theta) = \Psi_n(\omega t)$, which further leads to $\tilde{\Psi}_n(\omega t + \Delta\theta) = \tilde{\Psi}_{n+1}(\omega t)$. We translate the lattice position of the nonlinear wave function $\tilde{\Psi}_n(\omega t)$ by $N$ times, to find $\tilde{\Psi}_n(\omega t + N\Delta\theta) = \tilde{\Psi}_{n+N}(\omega t) \overset{\mathrm{PBC}}{=} \tilde{\Psi}_n(\omega t)$. This imposes the constraint $N\Delta\theta = 2m\pi$ with $m = 0, 1, 2, \ldots, N-1$. Finally, we dub $k = \Delta\theta = 2m\pi/N$ the wave number of the plane-wave nonlinear normal mode. It is at this point that we demonstrate the nonliner extension of Bloch theorem.

## IV. REFLECTION SYMMETRY AND QUANTIZED BERRY PHASE

This section aims to construct the nonlinear Berry phase of plane-wave nonlinear normal modes, reveal the reflection symmetry of the circuit dynamics, and demonstrate the topological invariance of nonlinear Berry phase.

Beginning from the plane-wave nonlinear normal mode in Eq.(S8), we realize the adiabatic evolution of the wavenumber $k(t')$ by allowing it to traverse through the Brillouin zone, from $k(0) = k$ to $k(t) = k + 2\pi$, while the amplitude $A$ of the nonlinear voltage fields in the circuit remains unchanged during this process. According to the nonlinear extension of the adiabatic theorem [5–8], a system $H(\Psi_k)$ initially in one of the nonlinear modes $\Psi_k$ will stay as an instantaneous nonlinear mode of $H(\Psi_{k(t)})$ throughout this procedure. Therefore, the only degree of freedom is the phase of mode. At time $t$, the mode is $\Psi_{k(t)}(\int_0^t \omega(t', q(t'))dt' - \gamma(t))$, where $\gamma(t)$ defines the phase shift of the plane-wave nonlinear normal mode in the adiabatic evolution. The dynamics of $\gamma$ follows from

$$\frac{d\gamma}{dt}\frac{\partial \Psi_k}{\partial \theta} = \frac{dk}{dt}\frac{\partial \Psi_k}{\partial k}. \tag{S9}$$

After $k$ traverses the Brillouin zone, the wave function acquires an extra phase $\gamma$, which we call Berry phase of nonlinear normal modes,

$$\gamma = \oint_{\mathrm{BZ}} dk \frac{\sum\limits_{l,j=1,2,3,4} \left( l \left| \psi_{l,k}^{(j)} \right|^2 \partial_k \phi_k^{(j)} + i \psi_{l,k}^{(j)*} \partial_k \psi_{l,k}^{(j)} \right)}{\sum\limits_{l',j'=1,2,3,4} l' \left| \psi_{l',k}^{(j')} \right|^2}. \tag{S10}$$

where $j, j' = 1, 2, 3, 4$ denote the four wave components of the classical field $\Psi_k$, and $\psi_{l,k}^{(j)} = (2\pi)^{-1} \int_0^{2\pi} e^{\mathrm{i}l\theta} \Psi_k^{(j)}$ is the $l$th Fourier component of $\Psi_k^{(j)}$. We note that in the linear limit, nonlinear Berry phase naturally reduces to the traditional format of linear Berry phase, $\gamma = \oint_{\mathrm{BZ}} dk\, \mathrm{i}\langle \Psi_k | \partial_k | \Psi_k \rangle$.

In a general nonlinear dynamics without any symmetry constraints, $\gamma$ is not quantized and can take any values between 0 and $\pi$. However, the underlying nonlinear circuit structure yields reflection symmetry, which subsequently quantizes the Berry phase of nonlinear normal modes. Below we first show that the system Lagrangian is invariant under reflection transformation, and then demonstrate the quantization of the nonlinear Berry phase.

The second-order time derivative of the voltage dynamics, or equivalently the four-field generalized nonlinear Schrödinger equations, are subjected to reflection symmetry. Under reflection transformation

$$\mathcal{M}(V_n^{(1)}, V_n^{(2)}) = (V_{-n}^{(2)}, V_{-n}^{(1)}), \tag{S11}$$

the system Lagrangian stays invariant via

$$\mathrm{L}(\{V_n^{(1)}, V_n^{(2)}\}) = \mathrm{L}(\{V_{-n}^{(2)}, V_{-n}^{(1)}\}). \tag{S12}$$

Thus, the Lagrangian equations of motion are also invariant under the reflection transformation. Similarly, the four-field generalized nonlinear Schrödinger equations also stay invariant under the reflection transformation of the classical field variable,

$$\mathcal{M}(\Psi_n^{(1)}, \Psi_n^{(2)}, \Psi_n^{(3)}, \Psi_n^{(4)}) = (\Psi_{-n}^{(3)}, \Psi_{-n}^{(4)}, \Psi_{-n}^{(1)}, \Psi_{-n}^{(2)}). \tag{S13}$$

Thus, given a plane-wave nonlinear normal mode $\Psi_k$, reflection symmetry guarantees a partner solution

$$\mathcal{M}\Psi_k = \\ (\Psi_k^{(3)}(\omega t + kn + \phi_k^{(3)}), \Psi_k^{(4)}(\omega t + kn + \phi_k^{(4)}), \Psi_k^{(1)}(\omega t + kn + \phi_k^{(1)}), \Psi_k^{(2)}(\omega t + kn + \phi_k^{(2)})) \tag{S14}$$

with the frequency $\omega$ and wave number $-k$. On the other hand, a plane-wave nonlinear normal mode of the wavenumber $-k$ and frequency $\omega$ is by definition denoted as

$$\Psi_{-k} = \\ (\Psi_{-k}^{(1)}(\omega t + kn + \phi_{-k}^{(1)}), \Psi_{-k}^{(2)}(\omega t + kn + \phi_{-k}^{(2)}), \Psi_{-k}^{(3)}(\omega t + kn + \phi_{-k}^{(3)}), \Psi_{-k}^{(4)}(\omega t + kn + \phi_{-k}^{(4)})) \tag{S15}$$

As there is no degeneracy of the nonlinear normal modes, $\Psi_{-k}$ and $\mathcal{M}\Psi_k$ have to be the same nonlinear mode, which in turn imposes the constraints

$$\phi_k^{(4)} - \phi_k^{(3)} = \phi_{-k}^{(2)} - \phi_{-k}^{(1)}, \qquad \phi_k^{(1)} - \phi_k^{(3)} = \phi_{-k}^{(3)} - \phi_{-k}^{(1)}, \tag{S16}$$

and

$$\Psi_k^{(1)} = \Psi_{-k}^{(3)}, \qquad \Psi_k^{(2)} = \Psi_{-k}^{(4)}. \tag{S17}$$

Eqs.(S16) and (S17) are the mathematical formulation of reflection symmetry in the nonlinear circuit dynamics.

Employing Eqs.(S16) and (S17) gives us the quantization of nonlinear Berry phase in Eq.(S10),

$$\gamma = \frac{1}{2}\oint_{\mathrm{BZ}} dk \frac{\partial}{\partial k}(\phi_k^{(3)} - \phi_k^{(1)}) = 0 \text{ or } \pi \mod 2\pi. \tag{S18}$$

In the last step of Eq.(S18), we employ the constraint from Eq.(S16): at time-reversal-invariant-momenta $k = 0, \pi$, both $\phi_\pi^{(3)} - \phi_\pi^{(1)}$ and $\phi_0^{(3)} - \phi_0^{(1)} = 0$ or $\pi$. It is at this point that we demonstrate the topological nature of the nonlinear Berry phase when reflection symmetry is considered. Furthermore, we note that the first and third components of $\Psi_n$ are the voltage fields, i.e., $\Psi_n^{(1)} = V_n^{(1)}$ and $\Psi_n^{(3)} = V_n^{(2)}$, and the second and fourth components of $\Psi_n$ are simply defined from the voltage fields via $\Psi_n^{(2)} = -\mathrm{i}\dot{V}_n^{(1)}$ and $\Psi_n^{(4)} = -\mathrm{i}\dot{V}_n^{(2)}$. Thus, the topological index can be equivalently expressed as the following form,

$$\gamma = \oint_{\mathrm{BZ}} dk \frac{\sum_l \left( l \left| v_{l,k}^{(j)} \right|^2 \partial_k \phi_k + \sum_{j=1,2} \mathrm{i} v_{l,k}^{(j)*} \partial_k v_{l,k}^{(j)} \right)}{\sum_{l',j'=1,2} l' \left| v_{l',k}^{(j')} \right|^2} = \phi_\pi - \phi_0 = n\pi, \tag{S19}$$

which is the topological number presented in the main text. Here, $\phi_k$ denotes the relative phase between the two components of plane-wave nonlinear voltage waves.

As indicated by Eq.(S19), the topological number can experience phase transition as $\phi_\pi$ and $\phi_0$ jump abruptly between 0 and $\pi$ for growing amplitudes. These relative phases are determined by comparing the frequencies $\omega(\phi_k = 0)$ and $\omega(\phi_k = \pi)$ for time-reversal-invariant-momenta $k = 0$ and $\pi$. $\gamma = \pi$ if $\omega(\phi_0 = 0)$ and $\omega(\phi_\pi = \pi)$ belong to the same nonlinear band, whereas $\gamma = 0$ if $\omega(\phi_0 = 0)$ and $\omega(\phi_\pi = 0)$ are in the same nonlinear band. $\gamma$ encounters a topological phase transition induced by the critical amplitude $A = A_c$ if the frequencies of upper and lower bands merge at

$$\omega(\phi_\pi = 0, A_c) = \omega(\phi_\pi = \pi, A_c). \tag{S20}$$

We perform the numerical analysis by computing the frequencies of the plane-wave nonlinear normal modes at time-reversal-invariant-momentum $k = \pi$ for a list of increasing amplitudes $A$. By comparing the frequencies of nonlinear normal modes with odd ($\phi_\pi = \pi$) and even ($\phi_\pi = 0$) parities, we use Eq.(S20) to numerically compute the topological phase transition amplitude $A_c$.

## V. THE NUMERICAL SIMULATION OF PROTOTYPE IN FIG.3 WITH ZERO RESISTANCE

In Fig.S3, the sharp and bright color at 18.86 MHz indicates that the response function on the open boundary is a delta-function, which stems from the standing bulk eigenmode. By comparing Fig.S3 with Fig.3(f-h), it is evident that the sharp response function is broadened into a Lorentzian form, whose half-width is proportional to the resistance. Consequently, the broadened response function leads to the "leakage" of standing bulk states into the band gap, and the leaked state has a spatial decay rate that is proportional to the resistance. This gives rise to the edge-like behavior that is observed in Fig.3(c-h), experimentally and numerically.

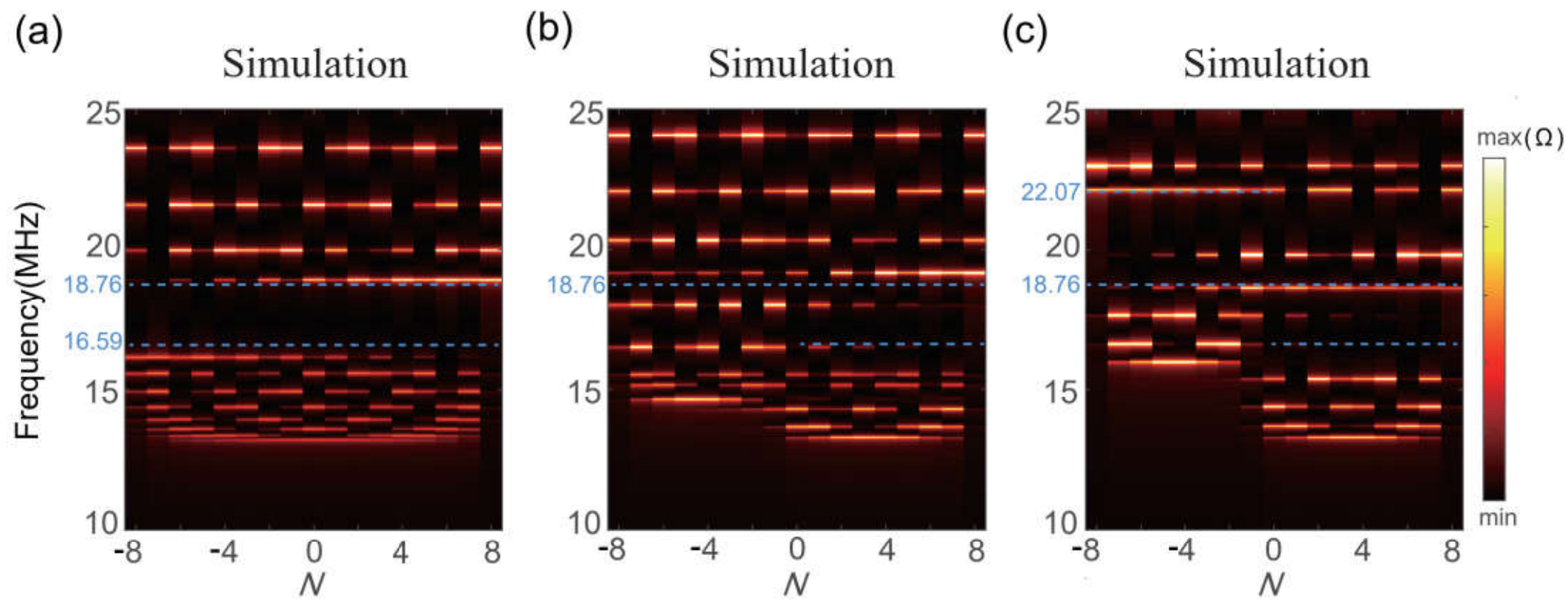

FIG. S3. Numerical simulations of the prototype in Fig.3(a), where the circuit resistance is set to zero. (a-c) Numerical simulations of Fig.3(c-e) respectively.

---

[1] Y. Wang, L.-J. Lango, C. H. Lee, B. Zhang, and Y.-D. Chong, Topologically enhanced harmonic generation in a nonlinear transmission line metamaterial, Nat. Commun. **10**, 1102 (2019).

[2] D. Zhou, D. Zeb Rocklin, M. Leamy, and Y. Yao, Topological Invariant and Anomalous Edge Modes of Strongly Nonlinear Systems, Nature Communications **13**, 3379 (2022).

[3] M. D. Fronk and M. J. Leamy, Higher-order dispersion, stability, and waveform invariance in nonlinear monoatomic and diatomic systems, Journal of Vibration and Acoustics **139** (2017).

[4] R. K. Narisetti, M. J. Leamy, and M. Ruzzene, A perturbation approach for predicting wave propagation in one-dimensional nonlinear periodic structures, Journal of Vibration and Acoustics **132** (2010).

[5] D. Xiao, M.-C. Chang, and Q. Niu, Berry phase effects on electronic properties, Rev. Mod. Phys. **82**, 1959 (2010).

[6] J. Liu, B. Wu, and Q. Niu, Nonlinear evolution of quantum states in the adiabatic regime, Phys. Rev. Lett. **90**, 170404 (2003).

[7] H. Pu, P. Maenner, W. Zhang, and H. Y. Ling, Adiabatic condition for nonlinear systems, Phys. Rev. Lett. **98**, 050406 (2007).

[8] J. Liu and L. B. Fu, Berry phase in nonlinear systems, Phys. Rev. A **81**, 052112 (2010).